\documentclass[aps,prb,preprint,superscriptaddress,showpacs]{revtex4-1}
\usepackage [dvips]{graphicx}
\usepackage{graphicx,color}
\usepackage{amsmath}
\usepackage{bm}
\usepackage{dcolumn}
\begin{document}

\title{Influence of the degree of decoupling of graphene\\
on the properties of transition metal adatoms}

\author{T.~Eelbo}
\affiliation{Institute of Applied Physics, University of Hamburg, Jungiusstr. 11, 20355 Hamburg, Germany}

\author{M.~Wa\'sniowska}
\email[Corresponding author: mwasniow@physnet.uni-hamburg.de]{}
\affiliation{Institute of Applied Physics, University of Hamburg, Jungiusstr. 11, 20355 Hamburg, Germany}

\author{M.~Gyamfi}
\affiliation{Institute of Applied Physics, University of Hamburg, Jungiusstr. 11, 20355 Hamburg, Germany}

\author{S.~Forti}
\affiliation{Max Planck Institute for Solid State Research, Heisenbergstr. 1, 70569 Stuttgart, Germany}

\author{U.~Starke}
\affiliation{Max Planck Institute for Solid State Research, Heisenbergstr. 1, 70569 Stuttgart, Germany}

\author{R.~Wiesendanger}
\affiliation{Institute of Applied Physics, University of Hamburg, Jungiusstr. 11, 20355 Hamburg, Germany}

\date{\today}

\begin{abstract}
We investigate the adsorption sites of $3d$ transition metal (TM) adatoms by means of low-temperature scanning tunneling microscopy and spectroscopy. Co and Ni adatoms were adsorbed on two types of graphene on SiC(0001), i.e. pristine epitaxial monolayer graphene (MLG) and quasi-free-standing monolayer graphene (QFMLG). In the case of QFMLG, two stable adsorption sites are identified, while in the case of MLG, only one adsorption site is observed. Our experimental results reveal the decoupling efficiency as a crucial parameter for determining the adsorption site as well as the electronic properties of $3d$ transition metal atoms on graphene. Furthermore, we show that Co atoms adsorbed on QFMLG are strong scattering potentials for Dirac fermions and cause intervalley scattering in their vicinity.
\end{abstract}


\maketitle

Graphene is a zero band gap semiconductor with the Dirac point (where the conduction and valence bands touch) located at the Fermi energy $E_{\rm{F}}$~\cite{Neto2009}. In the case of supported graphene, the position of the Dirac point (DP) with respect to ($E_{\rm{F}}$) can be affected by the interaction with the underlying substrate, the application of a gate voltage~\cite{Novoselov2004}, and chemical doping (e.g. by adatoms)~\cite{Schedin2007}. Tailor-made properties are a major aspect in current research~\cite{Riedl2010, Starke2012} and offer a possible route toward future graphene-based devices. However, the knowledge about the properties of adatoms on graphene is still limited. In particular, $3d$ transition metal (TM) adatoms so far have been mainly investigated by theory~\cite{Duffy1998, Yagi2004, Mao2008, Sevincli2008, Zanella2008, Johll2009,  Krasheninnikov2009, Cao2010, Valencia2010, Wehling2010, Sargolzaei2011, Uchoa2011, Wehling2011, Rudenko2012} and only few experimental studies have been performed~\cite{Brar2011, Gyamfi2011, Gyamfi2012}. Basic properties such as the adsorption site and the electronic configuration of $3d$ TM adatoms have not been studied yet.

From the viewpoint of theory, there are controversial predictions for $3d$ TM adatoms on graphene with respect to their electronic and magnetic properties. In density functional theory (DFT) calculations~\cite{Duffy1998, Yagi2004, Mao2008, Sevincli2008, Zanella2008, Krasheninnikov2009, Cao2010, Valencia2010, Sargolzaei2011, Uchoa2011} $3d$ TM atoms prefer to be adsorbed on the hollow site. By means of a generalized gradient approximation functional with an on-site Coulomb potential (GGA~+~$U$)~\cite{Wehling2010, Wehling2011} as well as in Ref.~\onlinecite{Johll2009}, top site adsorption was predicted. We note that the adsorption site strongly affects the calculated electronic and magnetic properties. In the present study, we pursued experimental investigations for two types of $3d$ TM atoms [Co and Ni (Ref.~\onlinecite{note1})] on two varieties of epitaxial graphene on SiC(0001). Different degrees of decoupling of the graphene layers from the substrate enable us to directly probe the influence of the properties of the graphene onto the adsorption sites of single atoms. The interaction with the underlying substrate or any corrugation within the graphene was neglected by theory so far. As we will show in this paper, the interaction of graphene with the substrate plays a crucial role for the properties of single adatoms, and, therefore, these investigations are of fundamental interest.

The graphene samples were prepared $ex-situ$ and precharacterized by angle resolved ultraviolet photoemission spectroscopy. The first type is pristine epitaxial monolayer graphene (MLG). It grows on top of a carbon buffer layer, often called zerolayer graphene (ZLG), and is n-doped due to the influence of the SiC-ZLG interface~\cite{Starke2009}. In contrast, the second type is a single carbon layer, namely ZLG, that was decoupled from the SiC substrate by hydrogen intercalation and, thus, turned into quasi-free-standing monolayer graphene (QFMLG)~\cite{Riedl2009, Forti2011}. The n-doping influence by the interface is removed and a slight p-doping develops~\cite{Forti2011}, which was proposed to be due to a surface charge layer in the substrate~\cite{Ristein2012}. Prior to low temperature scanning tunneling microscopy and spectroscopy (STM/STS) experiments ($T\approx5$~K), the samples were annealed ($\approx 800$~K) in ultrahigh vacuum to remove residual contaminations. Differential conductance (d$I$/d$U$) spectra were acquired using a lock-in technique with a modulation voltage of 10~mV and a frequency of 5~kHz. In STM topographies the adatom's position on graphene is indicated by means of the white lines, which are placed always on the hollow positions of the graphene lattice. In Fig.~\ref{Fig3}(b), the white lines indicate the top position of the graphene's unit cell. Co and Ni were evaporated at 12~K using e-beam evaporators.

\begin{figure}[t]
\begin{center}
\includegraphics[width=0.5\textwidth]{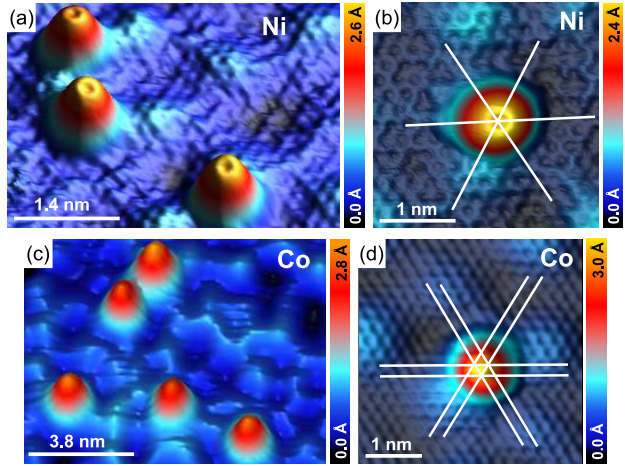}\\
\caption{(Color online)(a) and (c) STM topographies of MLG covered by Ni and Co adatoms, respectively ($U=0.15$~V, $I=0.5$~nA and $U=0.5$~V, $I=0.1$~nA). (b) and (d) The atomically resolved STM topographies indicating that Ni atoms adsorb on the hollow site and Co atoms on the top site on MLG ($U=-0.12$~V, $I=0.5$~nA and $U=-0.5$~V, $I=0.3$~nA).}\label{Fig1}
\end{center}
\end{figure}

We discuss the deposition of adatoms on MLG first, see Fig.~\ref{Fig1}. In the case of Ni we find all adatoms to be adsorbed in hollow sites [see Fig.~\ref{Fig1}(b)] in agreement with previous publications~\cite{Gyamfi2012, Eelbo2013}. The Ni adatoms exhibit a characteristic nodal structure which was related to a selective orbital coupling between Ni 3$d$ orbitals and graphene states. Due to symmetry considerations, this feature is exclusively observable for Ni adatoms adsorbed in hollow sites~\cite{Gyamfi2012}. Figure~\ref{Fig1}(c) shows Co decorated MLG. In contrast to Ni, we find top site adsorption for all Co adatoms, as depicted in Fig.~\ref{Fig1}(d) and in Ref.~\onlinecite{Eelbo2013}. The Co adatoms do not show a nodal structure.

\begin{figure}
\begin{center}
\includegraphics[width=0.5\textwidth]{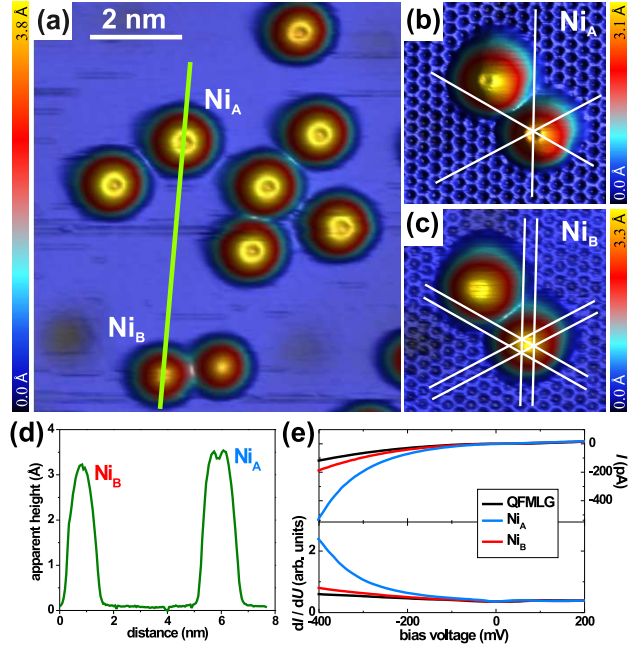}\\
\caption{(Color online) (a) STM image of QFMLG after the deposition of a small amount of Ni atoms at 12~K ($U=-0.5$~V and $I=0.1$~nA). (b) and (c) STM topographies revealing the adsorption geometry of Ni$_A$ and Ni$_B$ ($U=-0.1$~V, $I=0.09$~nA and $U=-0.05$~V, $I=0.09$~nA). The adsorption site of Ni atoms can be manipulated by the STM tip. (d) Line profile along the line indicated in (a) comparing Ni atoms located on two different adsorption sites. (e) STS spectra taken on Ni atoms adsorbed on QFMLG. Tunneling parameters for spectroscopic measurements are $U=0.4$~V and $I=0.05$~nA.}\label{Fig2}
\end{center}
\end{figure}

Similar experiments were performed for QFMLG as well. Figure~\ref{Fig2} shows QFMLG covered by Ni adatoms. The constant-current map as well as the line profiles show the presence of two different adsorbates. Ni$_{\rm{A}}$ exhibits an apparent height of 3.5~\AA~and a nodal structure similar to Ni/MLG. Not surprisingly, the adsorption site of Ni$_{\rm{A}}$ is the same, i.e. a hollow site, as shown in Fig.~\ref{Fig2}(b). Ni$_{\rm{B}}$ is adsorbed on the top site with an apparent height of 3.2~\AA~and, moreover, does not reveal a nodal structure, see Fig.~\ref{Fig2}(c). 

\begin{figure}[t]
\begin{center}
\includegraphics[width=0.5\textwidth]{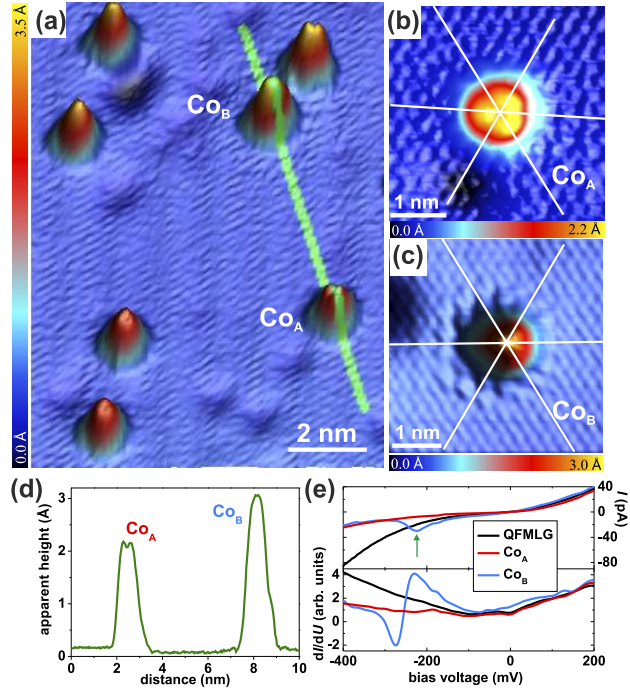}\\
\caption{(Color online) (a) Topographic image of two types of Co monomers on QFMLG ($U=0.3$~V and $I=0.1$~nA). (b) and (c) STM topographies indicating top site and hollow site adsorption in the case of Co$_{\rm{A}}$ and Co$_{\rm{B}}$, respectively ($U=-0.3$~V, $I=0.1$~nA and $U=0.3$~V, $I=0.1$~nA). (d) Line profile across Co$_{\rm{A}}$ and Co$_{\rm{B}}$ adatoms presented in (a). (e) Set of $I$ and d$I$/d$U$ spectra recorded on QFMLG and Co impurities. Tunneling parameters for STS are $I=0.07$~nA, $U=0.3$~V.}\label{Fig3}
\end{center}
\end{figure}

Co covered QFMLG is presented in Fig.~\ref{Fig3}(a). Similar to Ni/MLG, two types of Co monomers are found. The line profile in Fig.~\ref{Fig3}(d) shows different apparent heights and shapes of the adsorbates. Co$_{\rm{A}}$ exhibits an apparent height of 2.2~\AA, is adsorbed on a top site [see Fig.~\ref{Fig3}(b)], and is characterized by a "Y"--shaped depression on its center. The apparent height of Co$_{\rm{B}}$ is 3.1~\AA~while it exhibits a uniform shape and is found to adsorb on a hollow site, see Fig.~\ref{Fig3}(c). Furthermore, we investigated the adatoms by means of STS. In the case of  Co$_{\rm{B}}$, we found a change of the current at biases of about -0.25~V, which is marked by the green arrow in the upper panel of Fig.~\ref{Fig3}(e). This event goes hand in hand with a change of the shape and the adsorption site, i.e. Co$_{\rm{B}}$ changed to Co$_{\rm{A}}$. We note that in the case of Ni/QFMLG, such a switching process was never observed during STS, cf. Fig.~\ref{Fig2}(e).

The most striking difference between MLG and QFMLG concerning the adsorption sites of Ni and Co is the observation of either a single adsorption site for MLG or two adsorption sites for QFMLG. Since these results are independent of the deposited $3d$ transition metal adatoms, we have to relate the different adsorption sites to some of the different properties of QFMLG and MLG. The QFMLG surface is flat~\cite{Forti2011, Goler2013} while the MLG surface exhibits a commensurate superstructure of $(6\sqrt{3}\times6\sqrt{3})R30^{\circ}$ periodicity~\cite{Riedl2007}. At first glance, this corrugation might be considered to be the origin of the different adsorption. For this reason, we investigated the adsorption site of Co and Ni atoms within the $(6\sqrt{3}\times6\sqrt{3})R30^{\circ}$ unit cell. Ni and Co adatoms were always found to be adsorbed on the same lattice site, independent of the relative position within the unit cell of the superstructure of MLG. Moreover, Ni and Co adatoms preserve also the same electronic properties within the $(6\sqrt{3}\times6\sqrt{3})R30^{\circ}$ unit cell. This is in contrast to adatoms on the moir\'{e} pattern of graphene on Ru(0001)~\cite{Gyamfi2011} which exhibit different electronic properties in the case of being adsorbed on hills or valleys. Therefore, we consider the corrugation of MLG to be a negligible effect on the adsorption behavior and properties of atoms, even though in Ref.~\onlinecite{Valencia2010} theory predicts an influence of the curvature of graphene (e.g. nanotubes) on the adatoms' properties. The other major difference between QFMLG and MLG is the electronic structure, as for instance has been demonstrated by transport measurements~\cite{Jobst2010, Waldmann2011, Speck2011}. In case of QFMLG the topmost Si atoms are terminated by H atoms, so that the substrate should exhibit a large band gap similar to bulk SiC. In contrast, the electronic structure of MLG is strongly affected by the electronic states arising from the ZLG-SiC interface~\cite{Forti2011, Ohta2007}. This distinct difference between the substrates is the reason for the varying decoupling efficiencies, i.e. electronic properties, of MLG and QFMLG. Since the atomic structure of both types of graphene is identical, we relate the different adsorption sites of $3d$ TM atoms, i.e. one (two) adsorption site(s) on MLG (QFMLG), to these different electronic properties.

\begin{figure}[t]
\begin{center}
\includegraphics[width=0.5\textwidth]{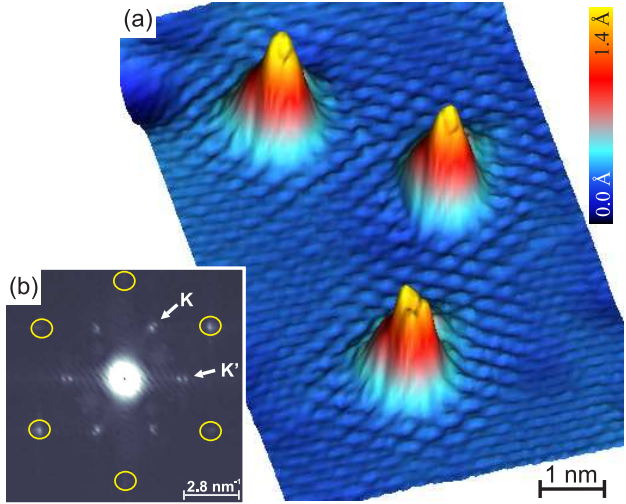}\\
\caption{(Color online) (a) Three dimensional STM image of QFMLG revealing a $(\sqrt{3}\times\sqrt{3})R30^{\circ}$ pattern in the vicinity of Co$_{\rm{A}}$ adatoms ($U=-0.3$~V and $I=70$~pA). (b) 2D-FFT analysis of the STM topography. Yellow circles indicate spots of the graphene lattice. The inner spots are related to intervalley scattering of electrons between nonequivalent Dirac cones. Its periodicity is $3.7$~\AA, equal to the wavelength $\lambda_{\rm F}=2 \pi / k_{\rm F}=3a/2$, where $a$ is the graphene lattice constant.}\label{Fig4}
\end{center}
\end{figure}

In the following we discuss our experimental results in view of already existing theoretical studies. For free-standing graphene Wehling et al.~\cite{Wehling2011} predicted different ground states of Co adatoms. Depending on the assumed on-site Coulomb potential ($U$), DFT calculations result in either hollow site adsorption with a $3d^94s^0$ configuration ($U=0$~eV) or top site adsorption with a $3d^84s^1$ configuration ($U=4$~eV). In contrast, Rudenko et al.~\cite{Rudenko2012} studied Co adatoms by means of a complete active space self-consistent field approach. This study found two stable configurations for Co adatoms located on the hollow site but at different heights. The corresponding electronic configurations are $3d^74s^2$ and $3d^94s^0$. Both theoretical models lack an explanation for the coexistence of hollow and top site adsorption on QFMLG. This sample system represents an efficiently decoupled graphene layer and is supposed to exhibit properties close to those of unsupported graphene  treated by theory. However, using $U=4$~eV Wehling et al. correctly predict the adsorption site for MLG. In case of Ni adatoms Wehling et al. obtained hollow site adsorption and a $3d^{10}4s^0$ configuration independent of the on-site Coulomb potential. These results fit well our experimental findings. The STS curves measured on Ni$_{\rm{A}}$ and Ni$_{\rm{B}}$ are similar; they do not show any peaks and only differ in the intensities of the d$I$/d$U$ signals due to different apparent heights of the adatoms in STM measurements. In the experiment we find differences of the apparent heights of about 0.3~\AA~for Ni and 0.9~\AA~for Co on QFMLG. The value for Ni is in perfect agreement with calculations~\cite{Wehling2011}. In contrast, the value for Co is about twice the calculated value. We relate this discrepancy to different electronic properties of Co$_{\rm{A}}$ and Co$_{\rm{B}}$ on QFMLG. 

So far, the discussion focused on the properties of $3d$ TM adatoms influenced by the properties of graphene. Another important aspect is the influence of the adatoms on the properties of graphene. Figure~\ref{Fig4}(a) reveals an interference pattern in the vicinity of several Co adatoms on QFMLG. This superstructure is confined to an area with a diameter of $\approx 3$~nm around the adatoms. The pattern is analyzed by means of a Fast Fourier analysis (FFT) of the STM topography as depicted in Fig.~\ref{Fig4}(b). The six outer spots, marked by yellow circles, originate from the hexagonal graphene lattice. The six inner spots correspond to a $(\sqrt{3}\times\sqrt{3})R30^{\circ}$ superstructure. Such a structure was observed before and was related to an intervalley scattering of the Dirac fermions of graphene~\cite{Rutter2007, Mallet2007}. In this process an electron is scattered between Dirac cones at two inequivalent $K$ points within the Brillouin zone ($K$ and $K'$). Such a scattering is induced and only allowed by a local symmetry breaking which manifests the impact of the adatoms on the graphene's electronic structure.

Interestingly, the intensity of the scattering pattern is lower for Co adatoms on hollow sites than on top sites, while the pattern is not visible for Ni adatoms on QFMLG independent of the adsorption site. Hence, we relate the scattering strength to the different electronic properties of the investigated adatoms. Ni was calculated to exhibit a $3d^{10}$ configuration independent of the adsorption site while a $3d^9$ and a $3d^8$ configuration was computed for Co adatoms on hollow and top sites, respectively~\cite{Wehling2011}. Thus, we suggest that the valence state of the adatoms has a strong impact on the strength of the scattering potential. We note, that in case of MLG, no intervalley scattering was observed for TM adatoms. Whether the absence of a scattering pattern in the constant-current maps is related to the degree of decoupling or to the corrugation superimposed by the carbon buffer layer could not be determined. 

In summary, we studied Co and Ni adatoms on two different model types of graphene on SiC(0001) as representatives for a stronger coupling to and decoupling from the substrate, respectively. We found that the adsorption geometry of Co and Ni atoms depends on graphene's electronic structures induced by the different degree of decoupling. In case of QFMLG the adatoms can be adsorbed on hollow and top sites of the graphene's lattice while in the case of MLG, only a single adsorption site is observed. These experimental findings disagree with recent theoretical predictions. Thus, theoretical investigations, which include the influence of the substrate, are highly desirable. Moreover, the experimental results indicate an adsorption-site-dependent electronic configuration of Co atoms while Ni atoms have always the same electronic configuration, i.e. predominantly $3d^{10}$. This is in agreement with calculations revealing different electronic configurations of Co adatoms for different adsorption sites~\cite{Wehling2011}. The different electronic properties explain the intensity variation of the intervalley scattering in the vicinity of adatoms on QFMLG as well. The $(\sqrt{3}\times\sqrt{3})R30^{\circ}$ interference pattern is the strongest in the case of Co atoms on top sites, while it is not observed in the case of Ni atoms with a fully occupied $d$-shell.

Financial support from the DFG (via SFB 668 and Grant No. WI 1277/25), the ERC Advanced Grant FURORE, Sta315/8-1 within the Priority Program 1459 Graphene and the Cluster of Excellence "Nanospintronics" is gratefully acknowledged.


\begin{thebibliography}{35}
\expandafter\ifx\csname natexlab\endcsname\relax\def\natexlab#1{#1}\fi
\expandafter\ifx\csname bibnamefont\endcsname\relax
  \def\bibnamefont#1{#1}\fi
\expandafter\ifx\csname bibfnamefont\endcsname\relax
  \def\bibfnamefont#1{#1}\fi
\expandafter\ifx\csname citenamefont\endcsname\relax
  \def\citenamefont#1{#1}\fi
\expandafter\ifx\csname url\endcsname\relax
  \def\url#1{\texttt{#1}}\fi
\expandafter\ifx\csname urlprefix\endcsname\relax\def\urlprefix{URL }\fi
\providecommand{\bibinfo}[2]{#2}
\providecommand{\eprint}[2][]{\url{#2}}

\bibitem[{\citenamefont{{Castro~Neto} et~al.}(2009)\citenamefont{{Castro~Neto},
  Guinea, Peres, Novosolev, and Geim}}]{Neto2009}
\bibinfo{author}{\bibfnamefont{A.~H.} \bibnamefont{{Castro~Neto}}},
  \bibinfo{author}{\bibfnamefont{F.}~\bibnamefont{Guinea}},
  \bibinfo{author}{\bibfnamefont{N.~M.~R.} \bibnamefont{Peres}},
  \bibinfo{author}{\bibfnamefont{K.~S.} \bibnamefont{Novosolev}},
  \bibnamefont{and} \bibinfo{author}{\bibfnamefont{A.~K.} \bibnamefont{Geim}},
  \bibinfo{journal}{Rev.~Mod.~Phys.} \textbf{\bibinfo{volume}{81}},
  \bibinfo{pages}{109} (\bibinfo{year}{2009}).

\bibitem[{\citenamefont{Novoselov et~al.}(2004)\citenamefont{Novoselov, Geim,
  Morozov, Jiang, Zhang, Dubonos, Grigorieva, and Firsov}}]{Novoselov2004}
\bibinfo{author}{\bibfnamefont{K.~S.} \bibnamefont{Novoselov}},
  \bibinfo{author}{\bibfnamefont{A.~K.} \bibnamefont{Geim}},
  \bibinfo{author}{\bibfnamefont{S.~V.} \bibnamefont{Morozov}},
  \bibinfo{author}{\bibfnamefont{D.}~\bibnamefont{Jiang}},
  \bibinfo{author}{\bibfnamefont{Y.}~\bibnamefont{Zhang}},
  \bibinfo{author}{\bibfnamefont{S.~V.} \bibnamefont{Dubonos}},
  \bibinfo{author}{\bibfnamefont{I.~V.} \bibnamefont{Grigorieva}},
  \bibnamefont{and} \bibinfo{author}{\bibfnamefont{A.~A.}
  \bibnamefont{Firsov}}, \bibinfo{journal}{Science}
  \textbf{\bibinfo{volume}{306}}, \bibinfo{pages}{666} (\bibinfo{year}{2004}).

\bibitem[{\citenamefont{Schedin et~al.}(2007)\citenamefont{Schedin, Geim,
  Morozov, Hill, Blake, Katsnelson, and Novoselov}}]{Schedin2007}
\bibinfo{author}{\bibfnamefont{F.}~\bibnamefont{Schedin}},
  \bibinfo{author}{\bibfnamefont{A.~K.} \bibnamefont{Geim}},
  \bibinfo{author}{\bibfnamefont{S.~V.} \bibnamefont{Morozov}},
  \bibinfo{author}{\bibfnamefont{E.~W.} \bibnamefont{Hill}},
  \bibinfo{author}{\bibfnamefont{P.}~\bibnamefont{Blake}},
  \bibinfo{author}{\bibfnamefont{M.~I.} \bibnamefont{Katsnelson}},
  \bibnamefont{and} \bibinfo{author}{\bibfnamefont{K.~S.}
  \bibnamefont{Novoselov}}, \bibinfo{journal}{Nat.~Mater.}
  \textbf{\bibinfo{volume}{6}}, \bibinfo{pages}{652} (\bibinfo{year}{2007}).

\bibitem[{\citenamefont{Riedl et~al.}(2010)\citenamefont{Riedl, Coletti, and
  Starke}}]{Riedl2010}
\bibinfo{author}{\bibfnamefont{C.}~\bibnamefont{Riedl}},
  \bibinfo{author}{\bibfnamefont{C.}~\bibnamefont{Coletti}}, \bibnamefont{and}
  \bibinfo{author}{\bibfnamefont{U.}~\bibnamefont{Starke}},
  \bibinfo{journal}{J.~Phys.~D:~Appl.~Phys.} \textbf{\bibinfo{volume}{43}},
  \bibinfo{pages}{374009} (\bibinfo{year}{2010}).

\bibitem[{\citenamefont{Starke et~al.}(2012)\citenamefont{Starke, Forti,
  Emtsev, and Coletti}}]{Starke2012}
\bibinfo{author}{\bibfnamefont{U.}~\bibnamefont{Starke}},
  \bibinfo{author}{\bibfnamefont{S.}~\bibnamefont{Forti}},
  \bibinfo{author}{\bibfnamefont{K.~V.} \bibnamefont{Emtsev}},
  \bibnamefont{and} \bibinfo{author}{\bibfnamefont{C.}~\bibnamefont{Coletti}},
  \bibinfo{journal}{MRS~Bull.} \textbf{\bibinfo{volume}{37}},
  \bibinfo{pages}{1177} (\bibinfo{year}{2012}).

\bibitem[{\citenamefont{Duffy and Blackman}(1998)}]{Duffy1998}
\bibinfo{author}{\bibfnamefont{D.~M.} \bibnamefont{Duffy}} \bibnamefont{and}
  \bibinfo{author}{\bibfnamefont{J.~A.} \bibnamefont{Blackman}},
  \bibinfo{journal}{Phys.~Rev.~B} \textbf{\bibinfo{volume}{58}},
  \bibinfo{pages}{7443} (\bibinfo{year}{1998}).

\bibitem[{\citenamefont{Yagi et~al.}(2004)\citenamefont{Yagi, Briere, Sluiter,
  Kumar, Farajian, and Kawazoe}}]{Yagi2004}
\bibinfo{author}{\bibfnamefont{Y.}~\bibnamefont{Yagi}},
  \bibinfo{author}{\bibfnamefont{T.~M.} \bibnamefont{Briere}},
  \bibinfo{author}{\bibfnamefont{M.~H.~F.} \bibnamefont{Sluiter}},
  \bibinfo{author}{\bibfnamefont{V.}~\bibnamefont{Kumar}},
  \bibinfo{author}{\bibfnamefont{A.~A.} \bibnamefont{Farajian}},
  \bibnamefont{and} \bibinfo{author}{\bibfnamefont{Y.}~\bibnamefont{Kawazoe}},
  \bibinfo{journal}{Phys.~Rev.~B} \textbf{\bibinfo{volume}{69}},
  \bibinfo{pages}{075414} (\bibinfo{year}{2004}).

\bibitem[{\citenamefont{Mao et~al.}(2008)\citenamefont{Mao, Yuan, and
  Zhong}}]{Mao2008}
\bibinfo{author}{\bibfnamefont{Y.}~\bibnamefont{Mao}},
  \bibinfo{author}{\bibfnamefont{J.}~\bibnamefont{Yuan}}, \bibnamefont{and}
  \bibinfo{author}{\bibfnamefont{J.}~\bibnamefont{Zhong}},
  \bibinfo{journal}{J.~Phys.:~Condens.~Matter} \textbf{\bibinfo{volume}{20}},
  \bibinfo{pages}{115209} (\bibinfo{year}{2008}).

\bibitem[{\citenamefont{{Sevin\c cli} et~al.}(2008)\citenamefont{{Sevin\c cli},
  Topsakal, Durgun, and Ciraci}}]{Sevincli2008}
\bibinfo{author}{\bibfnamefont{H.}~\bibnamefont{{Sevin\c cli}}},
  \bibinfo{author}{\bibfnamefont{M.}~\bibnamefont{Topsakal}},
  \bibinfo{author}{\bibfnamefont{E.}~\bibnamefont{Durgun}}, \bibnamefont{and}
  \bibinfo{author}{\bibfnamefont{S.}~\bibnamefont{Ciraci}},
  \bibinfo{journal}{Phys.~Rev.~B} \textbf{\bibinfo{volume}{77}},
  \bibinfo{pages}{195434} (\bibinfo{year}{2008}).

\bibitem[{\citenamefont{Zanella et~al.}(2008)\citenamefont{Zanella, Fagan,
  Mota, and Fazzio}}]{Zanella2008}
\bibinfo{author}{\bibfnamefont{I.}~\bibnamefont{Zanella}},
  \bibinfo{author}{\bibfnamefont{S.~B.} \bibnamefont{Fagan}},
  \bibinfo{author}{\bibfnamefont{R.}~\bibnamefont{Mota}}, \bibnamefont{and}
  \bibinfo{author}{\bibfnamefont{A.}~\bibnamefont{Fazzio}},
  \bibinfo{journal}{J.~Phys.~Chem.~C} \textbf{\bibinfo{volume}{112}},
  \bibinfo{pages}{9163} (\bibinfo{year}{2008}).

\bibitem[{\citenamefont{Johll et~al.}(2009)\citenamefont{Johll, Kang, and
  Tok}}]{Johll2009}
\bibinfo{author}{\bibfnamefont{H.}~\bibnamefont{Johll}},
  \bibinfo{author}{\bibfnamefont{H.~C.} \bibnamefont{Kang}}, \bibnamefont{and}
  \bibinfo{author}{\bibfnamefont{E.~S.} \bibnamefont{Tok}},
  \bibinfo{journal}{Phys.~Rev.~B} \textbf{\bibinfo{volume}{79}},
  \bibinfo{pages}{245416} (\bibinfo{year}{2009}).

\bibitem[{\citenamefont{Krasheninnikov
  et~al.}(2009)\citenamefont{Krasheninnikov, Lehtinen, Foster, Pyykk\"{o}, and
  Nieminen}}]{Krasheninnikov2009}
\bibinfo{author}{\bibfnamefont{A.~V.} \bibnamefont{Krasheninnikov}},
  \bibinfo{author}{\bibfnamefont{P.~O.} \bibnamefont{Lehtinen}},
  \bibinfo{author}{\bibfnamefont{A.~S.} \bibnamefont{Foster}},
  \bibinfo{author}{\bibfnamefont{P.}~\bibnamefont{Pyykk\"{o}}},
  \bibnamefont{and} \bibinfo{author}{\bibfnamefont{R.~M.}
  \bibnamefont{Nieminen}}, \bibinfo{journal}{Phys.~Rev.~Lett.}
  \textbf{\bibinfo{volume}{102}}, \bibinfo{pages}{126807}
  (\bibinfo{year}{2009}).

\bibitem[{\citenamefont{Cao et~al.}(2010)\citenamefont{Cao, Wu, Jiang, and
  Cheng}}]{Cao2010}
\bibinfo{author}{\bibfnamefont{C.}~\bibnamefont{Cao}},
  \bibinfo{author}{\bibfnamefont{M.}~\bibnamefont{Wu}},
  \bibinfo{author}{\bibfnamefont{J.}~\bibnamefont{Jiang}}, \bibnamefont{and}
  \bibinfo{author}{\bibfnamefont{H.-P.} \bibnamefont{Cheng}},
  \bibinfo{journal}{Phys. Rev. B} \textbf{\bibinfo{volume}{81}},
  \bibinfo{pages}{205424} (\bibinfo{year}{2010}).

\bibitem[{\citenamefont{Valencia et~al.}(2010)\citenamefont{Valencia, Gil, and
  Frapper}}]{Valencia2010}
\bibinfo{author}{\bibfnamefont{H.}~\bibnamefont{Valencia}},
  \bibinfo{author}{\bibfnamefont{A.}~\bibnamefont{Gil}}, \bibnamefont{and}
  \bibinfo{author}{\bibfnamefont{G.}~\bibnamefont{Frapper}},
  \bibinfo{journal}{J.~Phys.~Chem.~C} \textbf{\bibinfo{volume}{114}},
  \bibinfo{pages}{14141} (\bibinfo{year}{2010}).

\bibitem[{\citenamefont{Wehling et~al.}(2010)\citenamefont{Wehling, Balatsky,
  Katsnelson, Lichtenstein, and Rosch}}]{Wehling2010}
\bibinfo{author}{\bibfnamefont{T.~O.} \bibnamefont{Wehling}},
  \bibinfo{author}{\bibfnamefont{A.~V.} \bibnamefont{Balatsky}},
  \bibinfo{author}{\bibfnamefont{M.~I.} \bibnamefont{Katsnelson}},
  \bibinfo{author}{\bibfnamefont{A.~I.} \bibnamefont{Lichtenstein}},
  \bibnamefont{and} \bibinfo{author}{\bibfnamefont{A.}~\bibnamefont{Rosch}},
  \bibinfo{journal}{Phys.~Rev.~B} \textbf{\bibinfo{volume}{81}},
  \bibinfo{pages}{115427} (\bibinfo{year}{2010}).

\bibitem[{\citenamefont{Sargolzaei and Gudarzi}(2011)}]{Sargolzaei2011}
\bibinfo{author}{\bibfnamefont{M.}~\bibnamefont{Sargolzaei}} \bibnamefont{and}
  \bibinfo{author}{\bibfnamefont{F.}~\bibnamefont{Gudarzi}},
  \bibinfo{journal}{J.~Appl.~Phys.} \textbf{\bibinfo{volume}{110}},
  \bibinfo{pages}{064303} (\bibinfo{year}{2011}).

\bibitem[{\citenamefont{Uchoa et~al.}(2011)\citenamefont{Uchoa, Yang, Tsai,
  Peres, and {Castro Neto}}}]{Uchoa2011}
\bibinfo{author}{\bibfnamefont{B.}~\bibnamefont{Uchoa}},
  \bibinfo{author}{\bibfnamefont{L.}~\bibnamefont{Yang}},
  \bibinfo{author}{\bibfnamefont{S.-W.} \bibnamefont{Tsai}},
  \bibinfo{author}{\bibfnamefont{N.~M.~R.} \bibnamefont{Peres}},
  \bibnamefont{and} \bibinfo{author}{\bibfnamefont{A.~H.} \bibnamefont{{Castro
  Neto}}}, \bibinfo{journal}{arXiv:1105.4893v2}  (\bibinfo{year}{2011}).

\bibitem[{\citenamefont{Wehling et~al.}(2011)\citenamefont{Wehling,
  Lichtenstein, and Katsnelson}}]{Wehling2011}
\bibinfo{author}{\bibfnamefont{T.~O.} \bibnamefont{Wehling}},
  \bibinfo{author}{\bibfnamefont{A.~I.} \bibnamefont{Lichtenstein}},
  \bibnamefont{and} \bibinfo{author}{\bibfnamefont{M.~I.}
  \bibnamefont{Katsnelson}}, \bibinfo{journal}{Phys. Rev. B}
  \textbf{\bibinfo{volume}{84}}, \bibinfo{pages}{235110}
  (\bibinfo{year}{2011}).

\bibitem[{\citenamefont{Rudenko et~al.}(2012)\citenamefont{Rudenko, Keil,
  Katsnelson, and Lichtenstein}}]{Rudenko2012}
\bibinfo{author}{\bibfnamefont{A.~N.} \bibnamefont{Rudenko}},
  \bibinfo{author}{\bibfnamefont{F.~J.} \bibnamefont{Keil}},
  \bibinfo{author}{\bibfnamefont{M.~I.} \bibnamefont{Katsnelson}},
  \bibnamefont{and} \bibinfo{author}{\bibfnamefont{A.~I.}
  \bibnamefont{Lichtenstein}}, \bibinfo{journal}{Phys. Rev. B}
  \textbf{\bibinfo{volume}{86}}, \bibinfo{pages}{075422}
  (\bibinfo{year}{2012}).

\bibitem[{\citenamefont{Brar et~al.}(2011)\citenamefont{Brar, Decker, Solowan,
  Wang, Maserati, Chan, Lee, $\c{C}$. O.~Girit, Zettl, Louie
  et~al.}}]{Brar2011}
\bibinfo{author}{\bibfnamefont{V.~W.} \bibnamefont{Brar}},
  \bibinfo{author}{\bibfnamefont{R.}~\bibnamefont{Decker}},
  \bibinfo{author}{\bibfnamefont{H.-M.} \bibnamefont{Solowan}},
  \bibinfo{author}{\bibfnamefont{Y.}~\bibnamefont{Wang}},
  \bibinfo{author}{\bibfnamefont{L.}~\bibnamefont{Maserati}},
  \bibinfo{author}{\bibfnamefont{K.~T.} \bibnamefont{Chan}},
  \bibinfo{author}{\bibfnamefont{H.}~\bibnamefont{Lee}},
  \bibinfo{author}{\bibnamefont{$\c{C}$. O.~Girit}},
  \bibinfo{author}{\bibfnamefont{A.}~\bibnamefont{Zettl}},
  \bibinfo{author}{\bibfnamefont{S.~G.} \bibnamefont{Louie}},
  \bibnamefont{et~al.}, \bibinfo{journal}{Nat.~Phys.}
  \textbf{\bibinfo{volume}{7}}, \bibinfo{pages}{43} (\bibinfo{year}{2011}).

\bibitem[{\citenamefont{Gyamfi et~al.}(2011)\citenamefont{Gyamfi, Eelbo,
  Wa\'{s}niowska, and Wiesendanger}}]{Gyamfi2011}
\bibinfo{author}{\bibfnamefont{M.}~\bibnamefont{Gyamfi}},
  \bibinfo{author}{\bibfnamefont{T.}~\bibnamefont{Eelbo}},
  \bibinfo{author}{\bibfnamefont{M.}~\bibnamefont{Wa\'{s}niowska}},
  \bibnamefont{and}
  \bibinfo{author}{\bibfnamefont{R.}~\bibnamefont{Wiesendanger}},
  \bibinfo{journal}{Phys.~Rev.~B} \textbf{\bibinfo{volume}{83}},
  \bibinfo{pages}{153418} (\bibinfo{year}{2011}).

\bibitem[{\citenamefont{Gyamfi et~al.}(2012)\citenamefont{Gyamfi, Eelbo,
  Wa\'{s}niowska, Wehling, Forti, Starke, Lichtenstein, Katsnelson, and
  Wiesendanger}}]{Gyamfi2012}
\bibinfo{author}{\bibfnamefont{M.}~\bibnamefont{Gyamfi}},
  \bibinfo{author}{\bibfnamefont{T.}~\bibnamefont{Eelbo}},
  \bibinfo{author}{\bibfnamefont{M.}~\bibnamefont{Wa\'{s}niowska}},
  \bibinfo{author}{\bibfnamefont{T.~O.} \bibnamefont{Wehling}},
  \bibinfo{author}{\bibfnamefont{S.}~\bibnamefont{Forti}},
  \bibinfo{author}{\bibfnamefont{U.}~\bibnamefont{Starke}},
  \bibinfo{author}{\bibfnamefont{A.~I.} \bibnamefont{Lichtenstein}},
  \bibinfo{author}{\bibfnamefont{M.~I.} \bibnamefont{Katsnelson}},
  \bibnamefont{and}
  \bibinfo{author}{\bibfnamefont{R.}~\bibnamefont{Wiesendanger}},
  \bibinfo{journal}{Phys.~Rev.~B} \textbf{\bibinfo{volume}{85}},
  \bibinfo{pages}{161406(R)} (\bibinfo{year}{2012}).

\bibitem[{not()}]{note1}
\eprint{Co and Ni atoms were chosen as representatives of $3d$ TM atoms due to
  their high diffusion barriers.}

\bibitem[{\citenamefont{Starke and Riedl}(2009)}]{Starke2009}
\bibinfo{author}{\bibfnamefont{U.}~\bibnamefont{Starke}} \bibnamefont{and}
  \bibinfo{author}{\bibfnamefont{C.}~\bibnamefont{Riedl}},
  \bibinfo{journal}{J.~Phys.:~Condens.~Matter} \textbf{\bibinfo{volume}{21}},
  \bibinfo{pages}{134016} (\bibinfo{year}{2009}).

\bibitem[{\citenamefont{Riedl et~al.}(2009)\citenamefont{Riedl, Coletti,
  Iwasaki, Zakharov, and Starke}}]{Riedl2009}
\bibinfo{author}{\bibfnamefont{C.}~\bibnamefont{Riedl}},
  \bibinfo{author}{\bibfnamefont{C.}~\bibnamefont{Coletti}},
  \bibinfo{author}{\bibfnamefont{T.}~\bibnamefont{Iwasaki}},
  \bibinfo{author}{\bibfnamefont{A.~A.} \bibnamefont{Zakharov}},
  \bibnamefont{and} \bibinfo{author}{\bibfnamefont{U.}~\bibnamefont{Starke}},
  \bibinfo{journal}{Phys.~Rev.~Lett.} \textbf{\bibinfo{volume}{103}},
  \bibinfo{pages}{246804} (\bibinfo{year}{2009}).

\bibitem[{\citenamefont{Forti et~al.}(2011)\citenamefont{Forti, Emtsev,
  Coletti, Zakharov, Riedl, and Starke}}]{Forti2011}
\bibinfo{author}{\bibfnamefont{S.}~\bibnamefont{Forti}},
  \bibinfo{author}{\bibfnamefont{K.~V.} \bibnamefont{Emtsev}},
  \bibinfo{author}{\bibfnamefont{C.}~\bibnamefont{Coletti}},
  \bibinfo{author}{\bibfnamefont{A.~A.} \bibnamefont{Zakharov}},
  \bibinfo{author}{\bibfnamefont{C.}~\bibnamefont{Riedl}}, \bibnamefont{and}
  \bibinfo{author}{\bibfnamefont{U.}~\bibnamefont{Starke}},
  \bibinfo{journal}{Phys.~Rev.~B} \textbf{\bibinfo{volume}{84}},
  \bibinfo{pages}{125449} (\bibinfo{year}{2011}).

\bibitem[{\citenamefont{Ristein et~al.}(2012)\citenamefont{Ristein, Mammadov,
  and Seyller}}]{Ristein2012}
\bibinfo{author}{\bibfnamefont{J.}~\bibnamefont{Ristein}},
  \bibinfo{author}{\bibfnamefont{S.}~\bibnamefont{Mammadov}}, \bibnamefont{and}
  \bibinfo{author}{\bibfnamefont{T.}~\bibnamefont{Seyller}},
  \bibinfo{journal}{Phys.~Rev.~Lett.} \textbf{\bibinfo{volume}{108}},
  \bibinfo{pages}{246104} (\bibinfo{year}{2012}).

\bibitem[{\citenamefont{Eelbo et~al.}(2013)\citenamefont{Eelbo, Wa\'{s}niowska,
  Thakur, Gyamfi, Sachs, Wehling, Forti, Starke, Tieg, Lichtenstein,
  Wiesendanger et~al.}}]{Eelbo2013}
\bibinfo{author}{\bibfnamefont{T.}~\bibnamefont{Eelbo}},
  \bibinfo{author}{\bibfnamefont{M.}~\bibnamefont{Wa\'{s}niowska}},
  \bibinfo{author}{\bibfnamefont{P.}~\bibnamefont{Thakur}},
  \bibinfo{author}{\bibfnamefont{M.}~\bibnamefont{Gyamfi}},
  \bibinfo{author}{\bibfnamefont{B.}~\bibnamefont{Sachs}},
  \bibinfo{author}{\bibfnamefont{T.~O.}~\bibnamefont{Wehling}},
  \bibinfo{author}{\bibfnamefont{S.} \bibnamefont{Forti}},
  \bibinfo{author}{\bibfnamefont{U.}~\bibnamefont{Starke}},
  \bibinfo{author}{\bibfnamefont{C.}~\bibnamefont{Tieg}},
  \bibinfo{author}{\bibfnamefont{A.~I.}~\bibnamefont{Lichtenstein}},
  \bibnamefont{et~al.}, \bibinfo{journal}{Phys.~Rev.~Lett.}
  \textbf{\bibinfo{volume}{110}}, \bibinfo{pages}{136804} (\bibinfo{year}{2013}).
  
\bibitem[{\citenamefont{Goler et~al.}(2013)\citenamefont{Goler, Coletti,
  Piazza, Pingue, Colangelo, Pellegrini, Emtsev, Forti, Starke, Beltram
  et~al.}}]{Goler2013}
\bibinfo{author}{\bibfnamefont{S.}~\bibnamefont{Goler}},
  \bibinfo{author}{\bibfnamefont{C.}~\bibnamefont{Coletti}},
  \bibinfo{author}{\bibfnamefont{V.}~\bibnamefont{Piazza}},
  \bibinfo{author}{\bibfnamefont{P.}~\bibnamefont{Pingue}},
  \bibinfo{author}{\bibfnamefont{F.}~\bibnamefont{Colangelo}},
  \bibinfo{author}{\bibfnamefont{V.}~\bibnamefont{Pellegrini}},
  \bibinfo{author}{\bibfnamefont{K.~V.} \bibnamefont{Emtsev}},
  \bibinfo{author}{\bibfnamefont{S.}~\bibnamefont{Forti}},
  \bibinfo{author}{\bibfnamefont{U.}~\bibnamefont{Starke}},
  \bibinfo{author}{\bibfnamefont{F.}~\bibnamefont{Beltram}},
  \bibnamefont{et~al.}, \bibinfo{journal}{Carbon}
  \textbf{\bibinfo{volume}{51}}, \bibinfo{pages}{249} (\bibinfo{year}{2013}).

\bibitem[{\citenamefont{Riedl et~al.}(2007)\citenamefont{Riedl, Starke,
  Bernhardt, Franke, and Heinz}}]{Riedl2007}
\bibinfo{author}{\bibfnamefont{C.}~\bibnamefont{Riedl}},
  \bibinfo{author}{\bibfnamefont{U.}~\bibnamefont{Starke}},
  \bibinfo{author}{\bibfnamefont{J.}~\bibnamefont{Bernhardt}},
  \bibinfo{author}{\bibfnamefont{M.}~\bibnamefont{Franke}}, \bibnamefont{and}
  \bibinfo{author}{\bibfnamefont{K.}~\bibnamefont{Heinz}},
  \bibinfo{journal}{Phys.~Rev.~B} \textbf{\bibinfo{volume}{76}},
  \bibinfo{pages}{245406} (\bibinfo{year}{2007}).

\bibitem[{\citenamefont{Jobst et~al.}(2010)\citenamefont{Jobst, Waldmann,
  Speck, Hirner, Maude, Seyller, and Weber}}]{Jobst2010}
\bibinfo{author}{\bibfnamefont{J.}~\bibnamefont{Jobst}},
  \bibinfo{author}{\bibfnamefont{D.}~\bibnamefont{Waldmann}},
  \bibinfo{author}{\bibfnamefont{F.}~\bibnamefont{Speck}},
  \bibinfo{author}{\bibfnamefont{R.}~\bibnamefont{Hirner}},
  \bibinfo{author}{\bibfnamefont{D.~K.} \bibnamefont{Maude}},
  \bibinfo{author}{\bibfnamefont{T.}~\bibnamefont{Seyller}}, \bibnamefont{and}
  \bibinfo{author}{\bibfnamefont{H.~B.} \bibnamefont{Weber}},
  \bibinfo{journal}{Phys.~Rev.~B} \textbf{\bibinfo{volume}{81}},
  \bibinfo{pages}{195434} (\bibinfo{year}{2010}).

\bibitem[{\citenamefont{Waldmann et~al.}(2011)\citenamefont{Waldmann, Jobst,
  Speck, Seyller, Krieger, and Weber}}]{Waldmann2011}
\bibinfo{author}{\bibfnamefont{D.}~\bibnamefont{Waldmann}},
  \bibinfo{author}{\bibfnamefont{J.}~\bibnamefont{Jobst}},
  \bibinfo{author}{\bibfnamefont{F.}~\bibnamefont{Speck}},
  \bibinfo{author}{\bibfnamefont{T.}~\bibnamefont{Seyller}},
  \bibinfo{author}{\bibfnamefont{M.}~\bibnamefont{Krieger}}, \bibnamefont{and}
  \bibinfo{author}{\bibfnamefont{H.~B.} \bibnamefont{Weber}},
  \bibinfo{journal}{Nat.~Mater.} \textbf{\bibinfo{volume}{10}},
  \bibinfo{pages}{357} (\bibinfo{year}{2011}).

\bibitem[{\citenamefont{Speck et~al.}(2011)\citenamefont{Speck, Jobst, Fromm,
  Ostler, Waldmann, Hundhausen, Weber, and Seyller}}]{Speck2011}
\bibinfo{author}{\bibfnamefont{F.}~\bibnamefont{Speck}},
  \bibinfo{author}{\bibfnamefont{J.}~\bibnamefont{Jobst}},
  \bibinfo{author}{\bibfnamefont{F.}~\bibnamefont{Fromm}},
  \bibinfo{author}{\bibfnamefont{M.}~\bibnamefont{Ostler}},
  \bibinfo{author}{\bibfnamefont{D.}~\bibnamefont{Waldmann}},
  \bibinfo{author}{\bibfnamefont{M.}~\bibnamefont{Hundhausen}},
  \bibinfo{author}{\bibfnamefont{H.~B.} \bibnamefont{Weber}}, \bibnamefont{and}
  \bibinfo{author}{\bibfnamefont{T.}~\bibnamefont{Seyller}},
  \bibinfo{journal}{Appl.~Phys.~Lett.} \textbf{\bibinfo{volume}{99}},
  \bibinfo{pages}{122106} (\bibinfo{year}{2011}).

\bibitem[{\citenamefont{Ohta et~al.}(2007)\citenamefont{Ohta, Bostwick,
  McChesney, Seyller, Horn, and Rotenberg}}]{Ohta2007}
\bibinfo{author}{\bibfnamefont{T.}~\bibnamefont{Ohta}},
  \bibinfo{author}{\bibfnamefont{A.}~\bibnamefont{Bostwick}},
  \bibinfo{author}{\bibfnamefont{J.~L.} \bibnamefont{McChesney}},
  \bibinfo{author}{\bibfnamefont{T.}~\bibnamefont{Seyller}},
  \bibinfo{author}{\bibfnamefont{K.}~\bibnamefont{Horn}}, \bibnamefont{and}
  \bibinfo{author}{\bibfnamefont{E.}~\bibnamefont{Rotenberg}},
  \bibinfo{journal}{Phys.~Rev.~Lett.} \textbf{\bibinfo{volume}{98}},
  \bibinfo{pages}{206802} (\bibinfo{year}{2007}).

\bibitem[{\citenamefont{Rutter et~al.}(2007)\citenamefont{Rutter, Crain,
  Guisinger, Li, First, and Stroscio}}]{Rutter2007}
\bibinfo{author}{\bibfnamefont{G.~M.} \bibnamefont{Rutter}},
  \bibinfo{author}{\bibfnamefont{J.~N.} \bibnamefont{Crain}},
  \bibinfo{author}{\bibfnamefont{N.~P.} \bibnamefont{Guisinger}},
  \bibinfo{author}{\bibfnamefont{T.}~\bibnamefont{Li}},
  \bibinfo{author}{\bibfnamefont{P.~N.} \bibnamefont{First}}, \bibnamefont{and}
  \bibinfo{author}{\bibfnamefont{J.~A.} \bibnamefont{Stroscio}},
  \bibinfo{journal}{Science} \textbf{\bibinfo{volume}{317}},
  \bibinfo{pages}{219} (\bibinfo{year}{2007}).

\bibitem[{\citenamefont{Mallet et~al.}(2007)\citenamefont{Mallet, Varchon,
  Naud, Magaud, Berger, and Veuillen}}]{Mallet2007}
\bibinfo{author}{\bibfnamefont{P.}~\bibnamefont{Mallet}},
  \bibinfo{author}{\bibfnamefont{F.}~\bibnamefont{Varchon}},
  \bibinfo{author}{\bibfnamefont{C.}~\bibnamefont{Naud}},
  \bibinfo{author}{\bibfnamefont{L.}~\bibnamefont{Magaud}},
  \bibinfo{author}{\bibfnamefont{C.}~\bibnamefont{Berger}}, \bibnamefont{and}
  \bibinfo{author}{\bibfnamefont{J.-Y.} \bibnamefont{Veuillen}},
  \bibinfo{journal}{Phys.~Rev.~B} \textbf{\bibinfo{volume}{76}},
  \bibinfo{pages}{041403(R)} (\bibinfo{year}{2007}).

\end{thebibliography}
\end{document}